# mSCTP Based Decentralized Mobility Framework

[1]Waqas A. Imtiaz, [2]M. Afaq, [3]Mohammad A.U. Babar
[1,2,3]Department of Electrical Engineering
[1,3]IQRA National University, Peshawar, Pakistan
[2]City University of Science and Information Technology, Peshawar, Pakistan

*Abstract*— **To conceive the full potential of wireless IP services, Mobile Nodes (MNs) must be able to roam seamlessly across different networks. Mobile Stream Control Transmission Protocol (mSCTP) is a transport layer solution, which unlike Mobile IP (MIP), provides seamless mobility with minimum delay and negligible packet loss. However, mSCTP fails to locate the current IP address of the mobile node when Correspondent Node (CN) wants to initiate a session. In this paper, we propose DHT Chord to provide the required location management. Chord is a P2P algorithm, which can efficiently provide the IP address of the called MN by using its key-value mapping. The proposed decentralized mobility framework collectively exploits the multihoming feature of mSCTP, and efficient key-value mapping of chord to provide seamless mobility. Suitability of the framework is analyzed by preliminary analysis of chord lookup efficiency, and mSCTP handover procedure using overlay weaver and NS-2. Performance analysis shows that mSCTP multihoming feature and Chord efficient key-value mapping can provide a non-delayed, reliable, and an efficient IP handover solution.**

## I. INTRODUCTION

Basic design of Internet Protocol stack was laid on the assumption that, "*all nodes have fixed IP addresses*" [21]. This concept of fixed IP addresses worked flawlessly until communication nodes became mobile. Since then, IP mobility is a major issue, which needs to be resolved because when a MN changes its PoA, its IP address changes and results in termination of the ongoing session [1].

To exploit the full potential of wireless IP services, MNs must be able to wander seamlessly across different set of networks. Seamless mobility consists of two basic components: *handover management* and *location management*. Handover Management allows the MN to change its PoA without terminating the ongoing session. Location Management allows the MN to maintain its reachability for new connections after changing its PoA.

Layered Internet Protocol Stack enables us to provide seamless mobility at different layers [16]. MIP at the network layer provides a complete seamless mobility solution; however, its handover mechanism introduces unavoidable delay and dependence on additional network components [7]. A suitable handoff solution is required, which can provide mobility at user end without relying on additional network components, along with minimum delay and maximum security. SCTP is a transport layer protocol that encompasses revolutionary features like multihoming, multistreaming, and four-way handshake for connection establishment. SCTP multihoming feature can provide handover solutions; however, SCTP alone cannot support handoffs because it is not able to add or delete IP addresses during an active association. mSCTP, an extension of SCTP, enables SCTP to dynamically add/delete IP addresses during an active association and enables SCTP to perform handovers using its multihoming feature [1][7]. However, mSCTP fails to provide seamless mobility, when a CN wants to initiate a session with MN.

Location management in wireless IP services provides the IP address of the called MN. Traditional location management schemes like DDNS and SIP incorporates client server models, and suffers from their well-known drawbacks like congestion, centre point of failure and bottlenecking [16]. Such scheme is required, which along with mSCTP can provide an abrupt, decentralized, and reliable location management. DHT chord, a P2P algorithm, can provide the required location management by using its efficient key-value mapping as name to IP mapping.

Chord is a decentralized lookup system, which provides one and only operation: *given a key and it efficiently maps the key onto a peer* [1][7]. Chord forms a one-dimensional identifier circle, consisting of nodes and keys placed inside them, which ranges from 0 to $2^m$ - 1, where $m$ is the number of bits in the identifier circle. Each node and key is assigned an *m-bit* identifier; in order place them efficiently in the overlay network. $0(Log\ N)$ messages are required for key-value mapping over the identifier circle, where $N$ represents the number of nodes [12]. A complete decentralized approach, unnecessary complexity and no need for advanced technological changes, led us to investigate the performance and suitability of DHT Chord as location manager.

Efficient key-value mapping of chord can provide the necessary location management. Each node uses an identifier-locator set in the overlay network such that identifier refers to key i.e. Node ID, and locator represents the value or IP address in the key-value pair. Efficiency of the framework is increased via successor pointers, which efficiently reduces the number of messages during a chord query process. This paper proposes a complete decentralized mobility framework without unnecessary complexity and no need for advanced technological changes in the current internet architecture. Proposed framework exploits the multihoming feature of mSCTP for handover management at the transport layers, and efficient key-value mapping of chord for location management at the application layer to provide an efficient, robust and scalable mobility solution.

## II. RELATED WORK

In the near future, LTE and 4G technologies will not be able to support network-controlled handovers [47], therefore, an efficient user centric approach for seamless mobility is required. Kim and koh [5] evaluates the performance of MIP and mSCTP over IPv6 networks. Their analysis shows that mSCTP performs better than MIP via its multihoming feature. Zeadally and Siddiqui [10] show that mSCTP performs better than MIP and SIP in terms of handover latency and packet transmission after hadoffs. Ferrus and Brunstrom in [6], figures that transport layer is a worthwhile approach for handoffs and hence deserves more attention than the existing network and application layer solutions.

Park and kim [22] determines the performance of SCTP along with Session Initiation Protocol (SIP), in order to improve the Quality of Service (QoS) for real-time media. Their evaluation shows that mSCTP can perform better than UDP. Fu and Atiqquzama in [16] developes an analytical model to evalutae the performance of DNS as location manager. SCTP Draft proposes the use of Mobile IP for location managenet along with mSCTP, however this approach brings unavoidable delay via additional components.

P2P can perform lookups in a fraction of seconds with distributed content placement and discovery [11]. Cirani and Veltri in [14] propses an architecture for Distributed Location Service (DLS) which provides efficient location management. Sethom and Afifi in [13] presents PALMA (peer to peer architecture for location management), using tapestry for LM in mobile networks. However, PALMA cannot support handover management. Kunzmann and Hanks in [19] introduces a noval architecture for Next Generation Internet that is completely decentralized and relies on DHT algorithm.

## III. DECENTRALIZED MOBILITY FRAMEWORK

### A. Addressing Scheme:

An addressing scheme is required to identify and place mobile node when it enters the overlay network. Our addressing scheme consists of an *identifier-locator* set where identifier corresponds to the unique identity of the MN, and locator represents the current IP address in the network.

Each node inside the network has a *UID (Unique Identifier)* [1][14]. UID consists of three basic components, **name: device: ID**. Name contains the owner's name; it can be selected in any desired form, i.e. surname or name initials. Device refers to the type of the device, e.g. laptop, mobile, and PDA etc. ID is a unique identity that can be user's mobile number, email address or NIC number, e.g. *xyz: laptop: 17301xxxx*. Any naming scheme can be adapted as chord provides flexible naming mechanism.

Locator is referred to as *TL (Temporary Locator)* [1], which represents the value in the key-value pair. It is identical to the current IP address of the MN. TL has the ability to update itself as soon as MN attains a new IP address.

### B. Node Entry:

Mobile node must be able to join and publish its TL value as soon as it enters the network. When mobile node joins the network, it receives a new IP address and updates its UID-TL pair. After updating its pair, mobile node submits a query for mapping between its UID and base node in order to publish its TL value. Mapping process enables the mobile node to locate its corresponding base node, which will carry its TL value.

MN can update its UID-TL pair periodically by sending update messages to the base node. BN replies with an acknowledgment message. ACK messages are necessary to indicate the presence of the base node, as it is also mobile and can enter or leave the network at any instant. In case of acknowledgment failure, MN needs to locate another BN to publish its UID-TL value [1].

On reception of UID-TL value from the MN, base node publishes pointers towards its successor nodes. These pointers can shorten the query process by reducing the number of messages during a query. Successor pointers have the ability to time out if they are not periodically updated by their respective BN. Fig. 1 shows the process, where BN N8 publishes pointers towards its successors N12, N16 and N28. Base node N8 periodically updates these pointers information which helps in coping with any updated information regarding TL values [1].

### C. Location Search:

When CN wants to start a session, it requires MN's IP address, i.e. TL value. To obtain the current TL value of the MN, CN simply hashes its UID value and submits a query to its successor node. Successor node of the CN maps the given UID to the responsible base node and provides it with an IP address of the querying CN. This IP address helps to transfer the requested TL values directly to the CN.

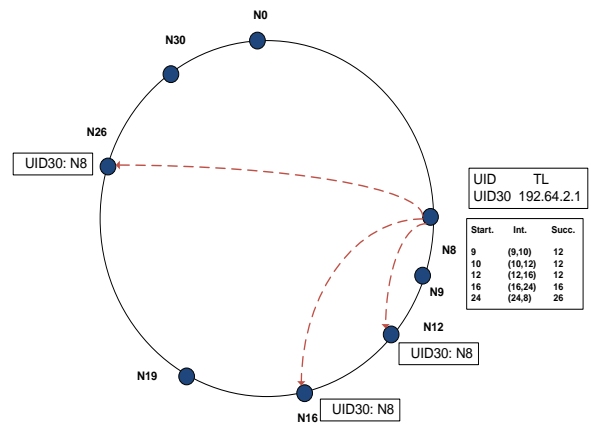

Figure 1: N8 publishing pointers towards its successor nodes

Pointers stored at every successor node shorten the number of $O(Log\ N)$ messages required to locate an object. For instance, the successor node encounters a node with a pointer towards the base node. In this case, the conventional query process terminates, and the query instantaneously redirects towards the base node, which provides the necessary TL value. Fig. 2, shows node N22 querying its successor node N26 for node N30 TL value. N22 first submits a query to its successor node N26. N26 having a pointer for UID30, redirects the query to N8 instead of searching for the closest interval in its finger table. Node N8 takes the required IP address of the querying node and directly transfers TL value of N30 to node N22. Thus, the use of successor pointers can reduce the number of query messages and hence the time required to find the required UID-TL pair.

### D. Location Update:

As mobile moves into a different network, its TL value and corresponding base node changes, this may involve the update of mobile node's location information. When MN enters the overlapping region, it attains a new IP address and updates its TL value. MN also sends an update message to its corresponding BN1 containing both TL values from network1 and network2 via interface 1. However, MN is still in AP1, and uses interface 1 as a primary path for communication.

When MN switches to network 2 in the overlapping region, it finds a new BN i.e. BN2 and updates its location with BN2 via interface 2, which is now the primary path for communication. Interface 1 is used as secondary path for redundancy purposes. MN after joining network2 informs BN2 of its previous base node, i.e. BN1. BN2 on receiving the necessary information sends a redirect query message towards BN1 [1]. This enables BN1 to redirect all the queries regarding UID value of the MN towards BN2, until its UID-TL value, and pointers time out in BN1 as shown in Fig. 3.

When MN leaves the overlapping region and completely transfers to network2, it updates its UID-TL value by sending an update message to the responsible BN2 via interface 2. TL value on interface 1 is dropped, so that the interface is free for further communication sessions.

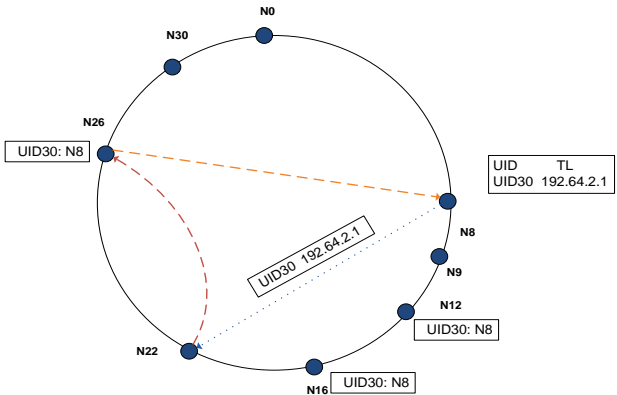

Figure 2: Query Process Made Easy with Successor Pointers

## IV. HANDOVER PROCEDURE

### A. Handover Procedure:

mSCTP decentralized mobility framework mainly undertakes the sessions originated from CN towards MN. DHT Chord in this scenario provides the current TL value of the MN, which initiates the session from CN towards the MN. Once the session is established, an mSCTP handover procedure supports the on-going session and handovers when required [8].

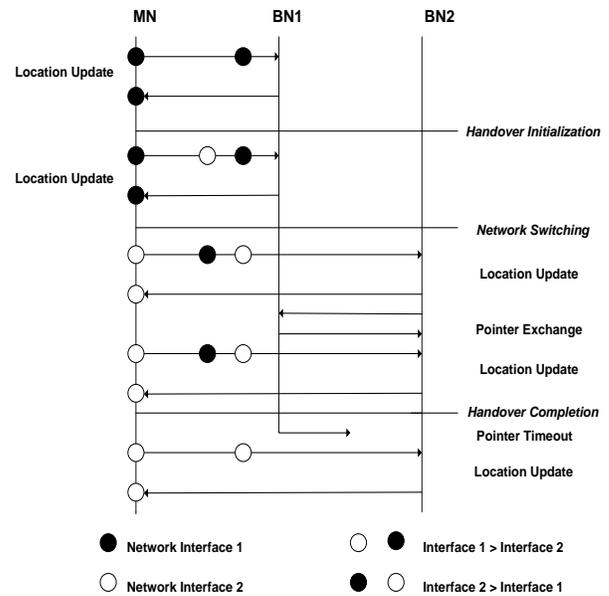

Figure 3: Location Update Process for MN [1]

### 1) Association Establishment:

In order to start a new session, CN simply hashes the UID value of a MN and submits a query to its closest successor. CN informs its mSCTP stack and initiates the basic SCTP association initialization process after receiving the called TL value. CN sends an INIT chunk towards the MN on the obtained TL (IP address) value. MN replies to INIT chunk with an INIT-ACK chunk, followed by connection establishment with the exchange of COOKIE-ECHO and COOKIE-ACK messages.

Fig. 4 shows an example where, MN N18 enters a new network i.e. network 1, and publishes its UID-TL pair via base node N24. On reception of the TL value, BN publishes the respective pointers towards its successor nodes. Now CN N2 wants to initiate session with the MN N18. CN first hashes UID of N18 and submits a query for its TL value. Node N24 provides the required TL value, which enables the CN to start a session with MN by the exchange of INIT, INIT-ACK, COOKIE-ECHO and COOKIE-ACK chunks between N2 and N18 between MN and CN.

### 2) Data Transport and Handover:

After association establishment, CN starts sending packets towards MN over the acquired TL value. BN contains the current location information of the MN, and periodically updates CN. MN periodically updates it TL value with its corresponding BN via update messages. CN also keeps itself updated by periodically querying its respective BN. Base node updates its successor nodes by periodically publishing pointers towards them, along with ACK replies to the MN.

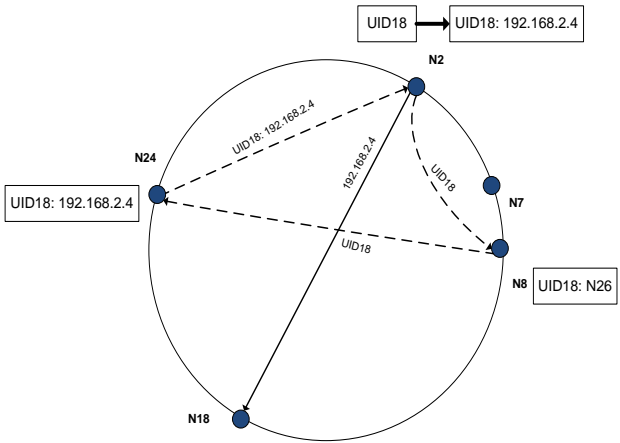

Figure 4: Association Establishment process using Chord as Location Manager

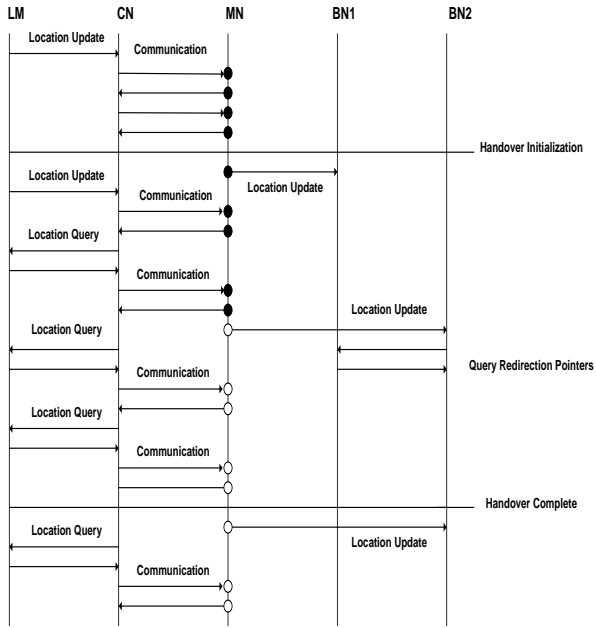

Figure 5: Data Transfer and Handover Procedure

Fig. 5 shows the data transport and handover process followed by MN and CN. LM handles any changes in MN's TL value and periodically updates CN about the BNs carrying the corresponding MN's TL value. Figure shows that CN continuously investigates location manager in order to keep itself updated of the BNs and MN TL value. When MN moves from network 1 towards network 2, it updates its respective BN i.e. BN2. BN1 after receiving query redirect message from BN2 updates its CN to locate BN2 for MN current TL value [1]. MN on the other hand performs a soft handover by adding the newly attained TL value to its ongoing SCTP association using mSCTP DAR extension.

## V. PRELIMINARY ANALYSIS

### A. mSCTP Handover Latency

#### 1) Analytical Analysis:

"Handover latency is the time at which an existing IP address becomes unavailable for end to end data transmission by movement of a node into a new network, to the time at which the end node receives a sequence of end to end transmission using the newly obtained IP address" [7].

$T_{\mathrm{mSCTP}}$ = mSCTP Handover Delay

$T_{\mathrm{md}}$ = Movement Detection

$T_{\mathrm{ac}}$ = Address Configuration

$T_{\mathrm{DAR}}$ = Dynamic Address Configuration Time

$T_{\mathrm{add\text{-}IP}}$ = Add-IP Address Chunk Exchange Time

$T_{\mathrm{pc\text{-}IP}}$ = Primary-IP Address Change Chunk Exchange Time

$T_{\mathrm{del\text{-}IP}}$ = Delete-IP address Chunk Exchange Time

$T_{\mathrm{pc}}$ = Primary Interface Change Time

mSCTP handover delay is accounted by MN's movement detection, Dynamic Address Reconfiguration (DAR) between MN and the CN, and configuration of the newly attained IP address. Thus the handover latency of mSCTP becomes [1][5][7] :

$$T_{\mathrm{mSCTP}} = T_{\mathrm{md}} + T_{\mathrm{ac}} + T_{\mathrm{DAR}} \qquad (1)$$

DAR includes the exchange of ADD-IP, DELETE-IP and PRIMARY CHANGE-IP messages along the processing time required to transfer the process these messages, so $T_{\mathrm{DAR}}$ in eq. 1 becomes [1],

$$T_{\mathrm{DAR}} = T_{\mathrm{add\text{-}IP}} + T_{\mathrm{pc\text{-}IP}} + T_{\mathrm{del\text{-}IP}} + T_{\mathrm{pc}} \qquad (2)$$

$$T_{\mathrm{add\text{-}IP}} + T_{\mathrm{pc\text{-}IP}} + T_{\mathrm{del\text{-}IP}} + T_{\mathrm{pc}} = (T_{\mathrm{MN\text{-}CN}} + T_{\mathrm{CN\text{-}MN}}) + (T_{\mathrm{MN\text{-}CN}} + T_{\mathrm{CN\text{-}MN}}) + (T_{\mathrm{MN\text{-}CN}} + T_{\mathrm{CN\text{-}MN}}) + T_{\mathrm{pc}} \qquad (3)$$

$$T_{\mathrm{add\text{-}IP}} + T_{\mathrm{pc\text{-}IP}} + T_{\mathrm{del\text{-}IP}} + T_{\mathrm{pc}} = 3(T_{\mathrm{MN\text{-}CN}} + T_{\mathrm{CN\text{-}MN}}) + T_{\mathrm{pc}} \qquad (4)$$

As MN is a multihoming device, so SCTP can utilize the primary interface to transmit all data chunks. Secondary interface is used to configure the newly acquired IP address. Thus, the time taken during movement detection and address configuration ($T_{\mathrm{md}} + T_{\mathrm{ac}}$) can be neglected, because MN and CN can communicate during these processes without session termination.

Therefore, the total handover latency in Eq. 1 becomes [1]

$$T_{\mathrm{mSCTP}} \cong \underbrace{T_{\mathrm{md}} + T_{\mathrm{ac}}}_{(\cong 0 \quad \text{due} \quad \text{to}} + 3(T_{\mathrm{MN\text{-}CN}} + T_{\mathrm{CN\text{-}MN}}) + T_{\mathrm{pc}} \qquad (5)$$

$$T_{\mathrm{mSCTP}} \cong 3(T_{\mathrm{MN\text{-}CN}} + T_{\mathrm{CN\text{-}MN}}) + T_{\mathrm{pc}} \qquad (6)$$

As explained in [8] [9]; that ASCONF chunks can be transmitted by bundling them with data chunks. Therefore, delay introduced by exchanging DAR control chunks between CN and MN, $3(T_{\mathrm{MN\text{-}CN}} + T_{\mathrm{CN\text{-}MN}})$, can be neglected as no extra time is spent in transmitting these chunks from MN towards CN. Only delay significant enough is the processing time of a MN during DAR procedure, which mainly includes switching of data transmission from one interface to another.

So the total theoretical handover latency for mSCTP in Eq. 6 becomes

$$T_{\text{mSCTP}} \cong \underbrace{T_{\text{md}} + T_{\text{ac}} + 3(T_{\text{MN-CN}} + T_{\text{CN-MN}})}_{} + T_{\text{pc}} \qquad (7)$$

*(≅0 bundling of ACONF chunks with Data chunks)*

$$T_{\text{mSCTP}} \cong T_{\text{pc}} \qquad (8)$$

Eq. 8 shows that the handover latency of mSCTP only accounts for the processing time of the node that is required to switch between interfaces. Thus, mSCTP can efficiently provide the required seamless mobility with minimum handover delay.

### 2) Simulation scenario:

A simulation scenario is designed to analyze the performance of mSCTP handover by measuring its handover latency. Handover latency is determined by measuring the time when the handover takes place to the time when a new packet arrives at the CN from the newly switched interface. Ns-2 uses SCTP multihoming feature to perform the required handover. Processes like that of mobile node movement detection, address configuration etc. are not taken into account, as delay introduced by them is negligible as shown in Eq. 7. Simulation scenario is run for 60 sec, and handover mechanism is introduced at 30 sec. Handover occurs when MN switches data transmission from one interface to another along with APs. Fig. 6 shows the data transmitted from MN towards the CN during 60 sec. Data transmission curve shows that there is no significant packet loss or delay during the handover process as the bytes transmitted keeps on increasing. There is a slight bent in the curve at 30 sec, which can be accounted for the processing time required to switch between the interfaces as mentioned in Eq. 8.

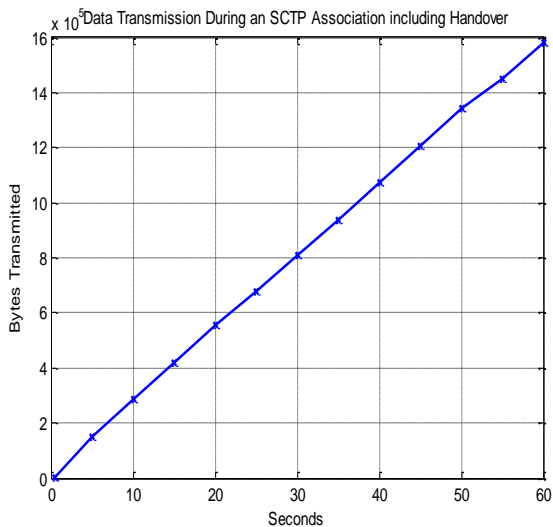

Figure 6: Data Transmitted During an SCTP Association Including Handover at 30 sec

### B. CHORD Successful Value Retrieval:

Suitability and success of chord as location manager depends on how efficiently it can retrieve the required key-value pair, and how often does it replies to queries requested by CN. Moreover chord must be able to self organize itself under worst network conditions. Overlay weaver is used to analyze the performance of chord, which is an overlay construction tool kit that supports multiple p2p lookup algorithms like CAN, Chord etc. It can invoke single or multiple nodes on the structured overlay network using multiple instances of DHT Shell. DHT shell is a layered command language interpreter that is used to control DHT and its algorithms. Each instance of DHT shell acts as a node on the overlay network.

### 1) Test Scenario:

To create a large overlay network in a rather small environment, we invoked several instances of DHT shell, over limited machines connected to one another in Local Area Network (LAN). The LAN consists of 8 computers with the following specifications, Intel Core 2 Duo 2 GHz processor and 3 GB RAM. Cisco 3600 series router and D-Link switch using Ethernet links connect these computers. Each node is running Windows XP and is equipped with Overlay weaver and Apache Ant build tool.

Each node is assigned a different port number to differentiate it from other nodes in the overlay network running on the same computer. 50 nodes of DHT shell are invoked on every computer, to get a large overlay network up to 400 nodes. Tests are performed manually, so no stabilization time or rate of nodes entering or leaving the network is considered. The ability of chord algorithm to support multiple values for a single key is used in our tests for performance analysis.

Successful retrieval of key-value pair is determined by making 25 queries in randomly selected nodes while increasing and decreasing the number of nodes. Four tests are conducted by varying the number of values associated with a key. Nodes are randomly selected to insert the key-value pair at the start of the test, using *Put <key> <value>*. To retrieve the required value in randomly selected nodes, *Get <value>* command is used at each instant. The number of queries successfully answered determines success of value retrieval. For decreasing network size, percentage of key-value retrieval is determined by decreasing the number of nodes in the overlay network from 400 to 100 nodes manually, and queries are made at random instances to retrieve the required key-value pair. The results obtained are explained below.

### 2) Results and Observations:

Results are observed as the percentage of queries successfully answered for a key-value i.e. UID-TL query. For this purpose, we first increased the number of nodes by introducing multiple DHT shells in each pc, and randomly made queries at different network sizes. Number of queries successfully answered are observed using DHT shell.

Fig. 7 shows that as we increase the number of nodes, amount of queries successfully answered somewhat decreases. However, the percentage of queries successfully answered still remains higher than 95%. Wrong/incomplete finger table entries and dropped UDP packets can be responsible for this decrease in successful retrieval of the key-value pair. Increase in the number of nodes, in the overlay network does not have any considerable effect on the performance of chord, as successful gets remains more than 95% in most cases.

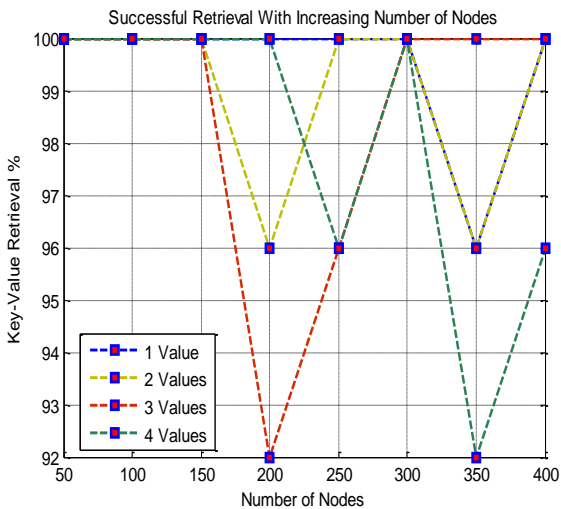

Figure 7: Successful Queries with Increasing Number of Nodes

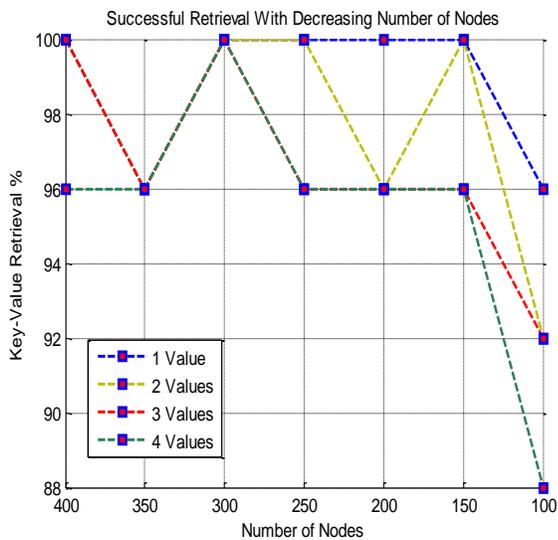

Figure 8: Successful Queries with Decreasing Network Size

When we decrease the number of nodes in the overlay network, value retrieval of chord remains efficient enough to operate as LM as shown in Fig. 8. In case of nodes, failures about 92 % of the queries are replied successfully. Even when the number of nodes is reduced from 400 to 100 nodes, chord is still able to retrieve the required key-value value. With decrease in the number nodes, the time taken to retrieve the required value increases, due to finger table updates or route failures. To account for this effect, we discarded any query that took more than 5 seconds. Performance analysis of chord value retrieval under both scenarios shows that chord can efficiently retrieve the required TL value, and hence can efficiently perform as location manager.

## VI. CONCLUSION

Seamless handoff solutions are required for future IP networks to facilitate ubiquitous users. Mobile Stream Control Transmission (mSCTP) protocol is a transport layer protocol, which enables MN to roam seamlessly across different networks. Unlike MIP, it provides handover solutions at user ends, with minimum delay and negligible packet loss via its multihoming feature. Preliminary analysis shows that mSCTP handover delay is as small as the processing time of the device, and it eliminates delay via agent discovery and registration process by exploiting its multihoming feature. However due to its inherent transport layer mechanism, mSCTP cannot provide location management.

Traditional location management schemes like DDNS and SIP incorporates client server models, which makes them vulnerable to centre point of failure, congestion and bottlenecking. DHT Chord can solve the problem by providing a complete decentralized approach using its efficient key-value mapping. An identifier-locator set is created using chord's key-value pair, which contains UID-TL pair to provide the required location management. Efficiency of the location management scheme is increased with the successor pointers. These pointers efficiently reduce the number of messages during query and hence the time in finding the required TL value. Performance analysis of chord value retrieval under both scenarios shows that, chord can efficiently retrieve the required TL value, and hence can efficiently perform as location manager.

This paper proposes a decentralized mobility framework for IP based handovers that does not requires any evolutionary technology changes to the current internet architecture. Efficient lookup algorithm, scalability, flexible naming, authorization support, and no central point of failure make DHT Chord an ideal candidate for location management. Supported by mSCTP multihoming feature and DAR extension, chord can efficiently provide the required name-IP mapping and support non-delayed handover procedures. Besides technical advantages, end users will gain added functionalities and more flexibility from this mobility framework.

## REFERENCES

[1] Waqas Ahmed, M. Jashim, "An SCTP Based Decentralized Mobility Framework", Masters Thesis, Department of Electrical Engineering, Blekinge Institute of Technlogy, Karlskrona, Sweden, July 2010.

[2] SCTP for Beginners, http://tdrwww.exp-math.uni-essen.de, [Online], [Accessed: Aug 5th, 2011].

[3] Muhammad Omer Chughtai, "Transport Protocol for Video Traffic over Mobile WiMAX", Master's Thesis, Department of Electrical Engineering, Comsats Institute of Information Technology, Islamabad, Pakistan, 2009.

[4] Randall Stewart, Paul D. Amer, Why is SCTP needed given TCP and UDP are widely available?, September 2007, http://www.isoc.org/briefings/017/, [Online], [Accessed: Aug 5th, 2011]

[5] Kim and S. Koh, "Analysis of handover latency for mobile IPv6 and mSCTP," *IEEE International Conference on Communication Workshops*, pp. 420-429, May 2008.

[6] L. Budzisz, R. Ferrus, A. Brunstrom, and F. Casadevall, "Towards transport-layer mobility: Evolution of SCTP multihoming," *Computer Communication*, vol. 31, pp. 980-998, Dec. 2008.

[7] J. K. Song, "Performance Evalution of Handoff between UMTS/802.11 Based on Mobile IP and Stream Control Transmission Protocol", Master's Thesis, Department of Computer Science, North Carolina State University, 2005.

[8] R. Stewart, Q. Xie, M. Tuexen, M. Maruyama, and M. Kozuka. (2007, Jun.) Internet-Draft SCTP Dynamic Address Reconfiguration. [Online]. http://tools.ietf.org/html/drafts/ietf-tsvwg-addip-sctp-22.

[9] S. J. Koh, Q. Xie, and S. D. Park. (2005, Oct.) Internet Draft mSCTP for IP Handover Support. [Online], http://www.ietf.org/proceedings/65/IDs?drafts-sjkoh-msctp-01.txt.

[10] S. Zeadally and F. Siddiqui, "An Empirical Analysis of Handoff Performance for SIP, Mobile IP, and SCTP Protocols," *Wireless personal communications*, vol. 43, pp. 589-603, 2007.

[11] I. Stoica, R. Morris, and D. Karger, "Chord: A scalable peer-to-peer lookup service for internet application," *ACM SIGCOMM Computer Communication*, vol. 31, pp. 149-160, 2001.

[12] S. Sarmady,"A survey on Peer-to-Peer and DHT", Grid Lab, School of Computer Science, Universiti Sains Malaysia, Penang, Malaysia, 2007.

[13] K. Sethom, H. Afifi, G. pujolle, "PALMA: A P2P based Architecture for Location Management", IFIP *International Conference on Mobile and Wireless Communications Networks,* 2005.

[14] S. Cirani, L. Veltri ,"Implementation of a Framework for DHT-based Distributed Location Service", *International Conference on Software, Telecommunication and Computer Networks*, p 279-83, 2008.

[15] M. Zangrilli, D. Bryan, "A Chord-based DHT for Resource Lookup in P2PSIP", Internet Dreaft draft-zangrilli-p2psip-dsip-dhtchord-00, SIPeerior Technologies, February 25, 2007.

[16] A. A. S. R. Reaz, M. Atiquzzaman, S. Fu, "Performance of DNS as location manager," *IEEE International Conference on Etectro Information Technology*, p. 6, May 2005.

[17] J. K. Song, "Performance Evalution of Handoff between UMTS/802.11 Based on Mobile IP and Stream Control Transmission Protocol", Master's Thesis, Department of Computer Science, North Carolina State University, 2005.

[18] S.Thomson, Y. Rekhter, Paul Vixie, "Dynamic Updates in the Domain Name System (DNS UPDATE)", RFC 2136

[19] O. Hanka, C. Spleib, G. Kunzmann, and J. Eberspächer, "A novel DHT-Based network architecture for the Next Generation Internet," *Eighth International Conference on Networks*, pp. 332-341, May 2009.

[20] F. Dabek, E. Brunskill, M. F. Kaashoek, and D. Karger, "Building peer-to-peer systems with chord, a distributed lookup service," *Proceeding of the eighth Workshop on operating System*, pp. 81-86, Aug. 2002.

[21] W. M. Eddy, "At What Layer Does Mobility Belong?," *IEEE Communication Magazine,*, vol. 42, no. 10, pp. 155-159, Oct. 2004.

[22] L. M. Campoy, " LTE-Advance Evolution to match Explosion of Mobile Data Traffic", 2[nd] International ICST Conference on Mobile Lightweight", Barcelone, Spain, 2010.

[23] H. Park, M. Kim, S. Lee, S. Kang, and Y. Kim, "A mobility management scheme using SCTP-SIP for real-time services across heterogeneous networks," *A mobility management scheme using SCTP-SIP for rea proceedingsof the 2009 ACM symposium on Applied Computing*, pp. 196-200, 2009.